**Oxygen Stoichiometry in Co-1212, Co-1222 and Co-1232 of Homologous Series Co-12*s*2 of "Category-B" Layered Copper Oxides**

Y. Morita, V.P.S. Awana, H. Yamauchi and M. Karppinen*

*Materials and Structures Laboratory, Tokyo Institute of Technology, 4259 Nagatsuta, Midori-ku, Yokohama 226-8503, Japan*

**Abstract**

Here results of a systematic study on oxygen stoichiometry are reported for the first three members of the novel $CoSr_2(Y,Ce)_sCu_2O_{5+2s\pm\delta}$ or Co-12*s*2 homologous series, *i.e.* Co-1212, Co-1222 and Co-1232 phases with a $SrO\text{-}CoO_{1\pm\delta}\text{-}SrO\text{-}CuO_2\text{-}(Y,Ce)\text{-}[O_2\text{-}(Y,Ce)]_{s\text{-}1}\text{-}CuO_2$ layered structure. The oxygen content was precisely determined by two independent chemical techniques: coulometric $Cu^+/Cu^{2+}$ titration and iodometric titration. Furthermore, oxygen stability/tunability was investigated by means of oxygenative and reductive annealings carried out in a thermobalance. It was revealed that all the three phases are rather stoichiometric in oxygen content and stable against both oxygenative and reductive annealings. The present results for the Co-12*s*2 homologous series suggest that not only the $CoO_{1\pm\delta}$ "charge reservoir" but also the nominally oxygen-stoichiometric $B\text{-}[O_2\text{-}B]_{s\text{-}1}$ "fluorite block", that is the characteristic structural element for the Co-12*s*2 ($s > 1$) phases and all other layered copper oxides of "category-B" [H. Yamauchi and M. Karppinen, *Superlatt. Microstructr.* **21A** (1997) 128] is non-tunable in terms of the oxygen content.

*Corresponding author

Prof. M. Karppinen

Materials and Structures Laboratory, Tokyo Institute of Technology, 4259 Nagatsuta, Midori-ku, Yokohama 226-8503, Japan

Phone: +81-45-924-5333

Fax: +81-45-924-5365

E-mail address: karppinen@msl.titech.ac.jp

## 1. Introduction

The ordered oxygen-vacancy perovskite block, $CuO_2$-($Q$-$CuO_2$)$_{n-1}$ ($Q$ = Ca or rare earth element), forms the basis for the $c$-axis-elongated multi-layered structures of high-$T_c$ superconductive and related copper oxides. When being piled together with other structural elements, *i.e.* other types of perovskite block, rock-salt layers and/or fluorite layers, it creates two categories of phases depending on the involvement of these structural elements: category-A includes the phases with perovskite and rock-salt layers of the sequence, $A$O-($M$O$_{1\pm\delta}$)$_m$-$A$O-$CuO_2$-($Q$-$CuO_2$)$_{n-1}$, while the structures of category-B contain fluorite layers as well according to the layer-sequence, $A$O-($M$O$_{1\pm\delta}$)$_m$-$A$O-$CuO_2$-$B$-($O_2$-$B$)$_{s-1}$-$CuO_2$. The stoichiometry of the phases thus obeys a general formula, $M_mA_rQ_{n-1}Cu_nO_{m+r+2n\pm\delta}$ (category-A) or $M_mA_{2k}B_sCu_{1+k}O_{m+4k+2s\pm\delta}$ (category-B), where $M$ = *e.g.* Cu, Bi, Pb, Tl, Hg, Al, C, and $A$ = *e.g.* Ba, Sr [1,2]. Each such compound is described in an unambiguous way with $M$-$m^{(A)}r^{(Q)}(n$-$1)n$ [$M$-$mr(n$-$1)n$] (category-A) or $M$-$m^{(A)}(2k)^{(B)}s(1+k)$ [$M$-$m(2k)s(1+k)$] (category-B). Furthermore, a group of category-A phases for which the $A$O-($M$O$_{1\pm\delta}$)$_m$-$A$O portion is common but the number of $CuO_2$ planes, $n$, varies comprises a homologous series, while in a homologous series of category-B, the members differ from one another in the number of fluorite layers, $s$, sandwiched by two $CuO_2$ planes [1,2].

The phases studied here, *i.e.* Co-1212 ($CoSr_2YCu_2O_{7\pm\delta}$), Co-1222 ($CoSr_2(Y_{3/4}Ce_{1/4})_2Cu_2O_{9\pm\delta}$) and Co-1232 ($CoSr_2(Y_{1/3}Ce_{2/3})_3Cu_2O_{11\pm\delta}$), possess structures described with a layer sequence of SrO-$CoO_{1\pm\delta}$-SrO-$CuO_2$-(Y,Ce)-[$O_2$-(Y,Ce)]$_{s-1}$-$CuO_2$ and obey the general chemical formula of $M_mA_{2k}B_sCu_{1+k}O_{m+4k+2s\pm\delta}$ or $M$-$m(2k)s(1+k)$ with $M$ = Co, $m$ = 1, $k$ = 1, and $s$ = 1, 2 and 3. The three phases thus together form a homologous series of category-B, *i.e.* Co-12$s$2. Here it should be noted that the Co-1212 phase may as well be categorized as an $m$ = 1, $r$ = 2, $n$ = 2 phase of category-A. Among the three phases, Co-1212 [3,4] and Co-1222 [5] have been reported earlier, but to the best of our knowledge no previous reports on Co-1232 exist.

As compared to category-A, members of category-B are rather poorly understood in terms of oxygen stoichiometry and doping. Here we apply two independent wet-chemical analysis methods, *i.e.* coulometric $Cu^+/Cu^{2+}$ titration and iodometric titration, to precisely establish the oxygen contents in all the three phases of the Co-12$s$2 series. In these redox methods applied, high-valent cation(s) of the sample, *i.e.* $Co^{III/IV}$ and/or $Cu^{III}$, are reduced by $Cu^+$ or $I^-$, respectively. As a second step in the analysis the exact amount of the left-over reductant ($Cu^+$) or the oxidized form of the reductant as formed in the redox reaction ($I_2$) is determined using an appropriate technique, *i.e.* anodic oxidation ($Cu^+$) or conventional titration ($I_2$) [6,7]. Furthermore, the stability/tunability of the oxygen stoichiometry is investigated for each phase by thermogravimetric (TG) ex-periments carried out under both reductive and oxygenative conditions.

## 2. Experimental

Samples of $CoSr_2YCu_2O_{7\pm\delta}$ (Co-1212), $CoSr_2(Y_{3/4}Ce_{1/4})_2Cu_2O_{9\pm\delta}$ (Co-1212) and $CoSr_2(Y_{1/3}Ce_{2/3})_3Cu_2O_{11\pm\delta}$ (Co-1232) were synthesized with a solid-state reaction method. The starting materials, $Co_2O_3$, $SrO_2$, $Y_2O_3$, $CeO_2$ and CuO, were mixed to appropriate ratios and calcined at 975 $^oC$ and 1000 $^oC$ for Co-1212 with an intermediate grinding. For Co-1222 and Co-1232, additional heat treatments at temperatures 10 to 20 $^oC$ higher than those for Co-1212 were necessary to obtain the desired phases. The calcined powders were pressed into pellets and annealed in flowing oxygen at 1010 $^oC$ for 40 h followed by a slow cooling down to room temperature. The phase purity of each sample was confirmed with a powder x-ray diffraction measurement (XRD; MAC Science: MXP18VAHF[22]; Cu $K_\alpha$ radiation). Magnetization measurements were carried out with a SQUID magnetometer (Quantum Design: MPMS-XL).

The precise values of oxygen content of the as-synthesized samples were determined with $Cu^+/Cu^{2+}$ coulometric titrations and iodometric titrations. Coulometric titration experiments were repeated several times to confirm the accuracy. Each titration experiment was performed in an air-tight cell under an Ar atmosphere. Coulometric titrations were carried out using CuCl to reduce high-valent Cu and Co to $Cu^{2+}$ and $Co^{2+}$ ions [7]. The sample (*ca*. 20 mg) was dissolved in 1 M HCl solution (*ca*. 200 ml) containing a presicely known excess of $Cu^+$ ions and the remaining excess of them was determined by means of electrochemical oxidation, *i.e.* coulometric titration. The anodic oxidation of $Cu^+$ to $Cu^{2+}$ was performed with a dc current of 5 mA. Platinum plates were used as electrodes and the cathode was separated from the cell solution with a salt bridge. The potential was measured against an Ag/AgCl reference electrode (Horiba pH Meter F-22). The end-point of the titration was determined at 930 mV. Before each titration, a pre-titration with a small amount of CuCl (2-3 mg) was done to standardize the starting potential of the cell solution. Several blank titrations were carried out with the CuCl powder to standardize the titration. In iodometric titrations the sample (*ca.* 40 mg) was dissolved in 1 M HCl solution (*ca.* 100 ml) containing an excess of KI (*ca.* 1 g), resulting in reduction of high-valent Cu and Co species of the sample to solid copper(I) iodide and $Co^{2+}$ ions [7], and formation of a stoichiometric amount of iodine in the solution. Iodine was titrated with $Na_2S_2O_3$ solution (*ca.* 0.015 M). The end-point was detected visually using starch as the indicator.

The possible changes in the oxygen stoichiometry of the as-synthesized samples upon oxygenative/reductive annealings were investigated by means of TG analysis carried out in a thermobalance of high sensitivity (Perkin-Elmer: TAS7). In these measurements a slow temperature scanning (1 $^{o}$C/min) from room temperature to 800 $^{o}$C and 950 $^{o}$C in Ar and $O_2$ gas flows, respectively, was applied. The sample mass was *ca.* 100 mg.

## 3. Results and discussion

From the x-ray diffraction patterns (Fig. 1) all the three samples of the Co-12*s*2 series were found of single phase. No sign of intergrowths of other Co-12*s*2 phases was seen. This is most clearly evidenced from the fact that each sample possessed a distinct low-angle 00l diffraction peak only. Earlier it was reported that for Co-1212 the basic *M*-1212 cell is doubled due to two distinct Co sites in alternate sub-cells, thus resulting in the space group *Ima*2 instead of *P*4/*mmm* [3,4]. Here we found that the situation is the same for the novel Co-1232 phase. On the other hand, the XRD pattern of Co-1222 could be indexed according to the space group *I*4/*mmm* [5]. Note that for Co-1222 the basic unit cell is body centered and hence further doubling of the cell is not required to accommodate the two different Co sites. The *c*-axis lattice parameter was determined for the three phases from the XRD data at 22.80, 28.17 and 33.23 Å for Co-1212, Co-1222 and Co-1232, respectively. For the sake of an easier comparison among the three phases, this was done considering only the unit cell doubling within the *P*4/*mmm* space group for Co-1212 and Co-1232. Indexing in the x-ray patterns of Fig. 1 was done on the same basis. Detailed high-resolution transmission electron microscopy studies on these compounds are reported elsewhere [8], where the change in space group from *P*4/*mmm* to *Ima*2 is verified and various superstructures are revealed for Co-1212 and Co-1232 due to different arrangements of the $CoO_4$ tetrahedra.

The oxygen contents determined for the as-synthesized Co-1212, Co-1222 and Co-1232 samples with the two independent wet-chemical analysis methods are given in Table 1. The two methods, *i.e.* $Cu^{+}/Cu^{2+}$ coulometric titration and iodometric titration, revealed essentially identical numbers for the oxygen contents for all the three samples. The oxygen contents as calculated from coulometric titration data were 6.99, 8.99 and 10.99 *per* formula unit for Co-1212, Co-1222 and Co-1232, respectively, while iodometric titration revealed oxygen contents of 6.97, 8.98 and 10.98 for the same samples. The presently obtained value of ~6.98 for the oxygen content of the Co-1212 sample is in good agreement with the value of ~7.01 calculated for the same sample from the oxygen-site occupancies refined from neutron diffraction data [4].

Thermogravimetric curves for the reductive Ar annealings carried out for the as-synthesized Co-1212, Co-1222 and Co-1232 samples are shown in Fig. 2. No weight change is seen in any of the curves. This behaviour is definitely different from that well established for the Cu-1212 ($CuBa_2YCu_2O_{7-\delta}$) phase. In other words, the “excess” oxygen on the $CoO_{1\pm\delta}$ ($\delta \approx 0$) layer of the Co-12*s*2 phases is stable against reductive heat treatments. (Needless to say, from the $CuO_{1-\delta}$ chain of the Cu-1212 phase it is easy to gradually remove oxygen by means of Ar annealing.) Here it should be noted that not only the Ar annealing but also the $O_2$ annealing yielded no changes in the oxygen content for any of the phases, Co-1212, Co-1222 or Co-1232. In the Co-12*s*2 structures, the excess-oxygen atoms are arranged such that they form chains of corner-sharing $CoO_4$ tetrahedra rather than planar $CoO_4$ squares [3-5]. Such a $CoO_4$ tetrahedron thus seems to be an extremely rigid configuration in terms of oxygen content.

The present results allow us to make another important conclusion: our two observations (*i*) that the compounds are stoichiometric in the overall oxygen content, and (*ii*) that no oxygen is lost upon reductive and oxygenative annealings, are considered as a manifestation of the fact that not only the $CoO_{1\pm\delta}$ layer but also the (Y,Ce)-$[O_2\text{-(Y,Ce)}]_{s-1}$ fluorite block in the Co-12*s*2 phases is oxygen stoichiometric and stable against reductive and oxygenative annealings. This further suggests that the $B$-$[O_2\text{-}B]_{s-1}$ fluorite block of category-B layered copper oxides is in general non-tunable in terms of oxygen content.

None of the three samples showed superconductivity. The dc magnetization data obtained for the three phases are reminiscent of a general common behaviour being characteristic to low-dimensional magnetic Co-O layers. The magnetic characteristics for Co-1212 were discussed elsewhere [4], and those for Co-1222 are plotted in Fig. 3. The ZFC (zero-field cooled) and FC (field-cooled) magnetization curves start branching around 50 K clearly, exhibiting small additional kinks in the ZFC branch. Since the magnetization *versus* field data in these temperature ranges do not exhibit any hysteresis loops (see the inset in Fig. 3) even though the ZFC and FC branching is indicative of some ferromagnetic component, one may assume that the magnetic behavior of the Co-O layer of Co-1222 is due to short-range correlations among the Co spins. The magnetization behaviour of Co-1232 was found to be similar to that of Co-1222. Detailed magnetization and neutron scattering studies are warranted to fully understand the magnetic interactions of Co spins in these compounds.

## 4. Conclusion

We showed that the three first members of the Co-12*s*2 homologous series of category-B are all essentially stoichiometric and non-tunable in terms of the oxygen content. Unconventional magnetic behaviours were observed common to all the three phases.

## Acknowledgment

The present work has been supported by a Grant-in-Aid for Scientific Research (contract No. 11305002) from the Ministry of Education, Science and Culture of Japan.

**Table 1.** Value of oxygen content *per* formula unit for the as-synthesized $CoSr_2YCu_2O_{7\pm\delta}$ (Co-1212), $CoSr_2(Y_{3/4}Ce_{1/4})_2Cu_2O_{9\pm\delta}$ (Co-1222) and $CoSr_2(Y_{1/3}Ce_{2/3})_3Cu_2O_{11\pm\delta}$ (Co-1232) samples as determined by $Cu^{+}/Cu^{2+}$ coulometric titrations and iodometric titrations in one to three parallel experiments.

| Sample | Analysis method | Exp. 1 | Exp. 2 | Exp. 3 | Average | |
|---|---|---|---|---|---|---|
| Co-1212 | Coulometry | 6.99 | 6.99 | 6.98 | 6.99 | 6.98 |
| | Iodometry | 6.97 | | | 6.97 | |
| Co-1222 | Coulometry | 8.99 | 8.97 | 9.02 | 8.99 | 8.99 |
| | Iodometry | 8.98 | | | 8.98 | |
| Co-1232 | Coulometry | 10.99 | 10.98 | | 10.99 | 10.99 |
| | Iodometry | 10.98 | | | 10.98 | |

**Figure captions**

**Fig.** 1. XRD patterns obtained for as-synthesized (a) Co-1212, (b) Co-1222, and (c) Co-1232 samples.

**Fig.** 3. TG curves recorded for as-synthesized (a) Co-1212, (b) Co-1222, and (c) Co-1232 samples in flowing Ar gas. The heating and cooling rate was 1 $^{o}C$/min and the mass of the sample *ca*. 100 mg.

**Fig.** 3. Magnetic susceptibility ($\chi$) *versus* temperature ($T$) plot for the as-synthesized Co-1222 sample. Inset shows the *M-H* behaviour for the same sample at 5 K.

PCP49 / Figure 1 (Morita *et al.*)

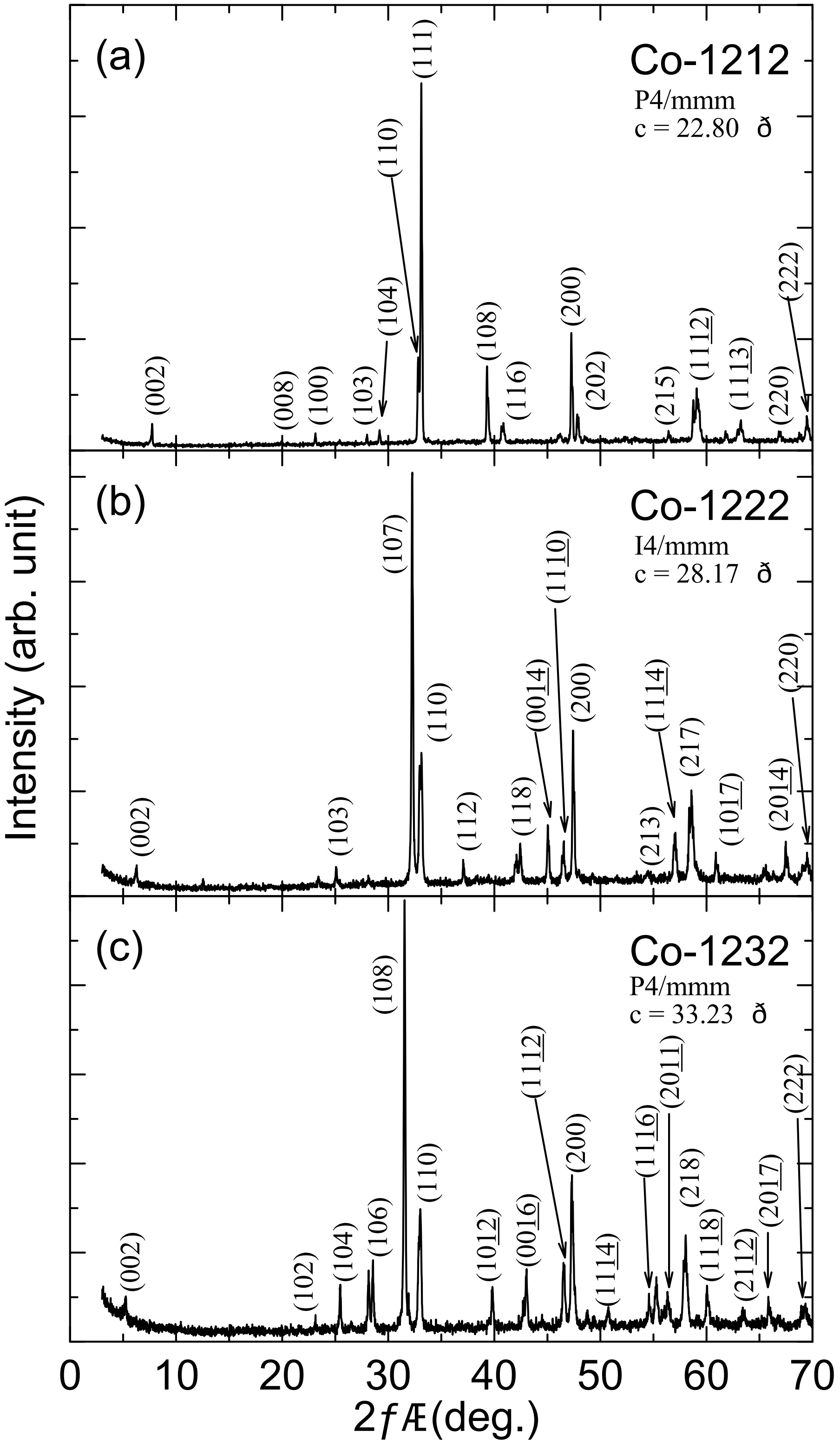

PCP49 / Figure 2 (Morita *et al.*)

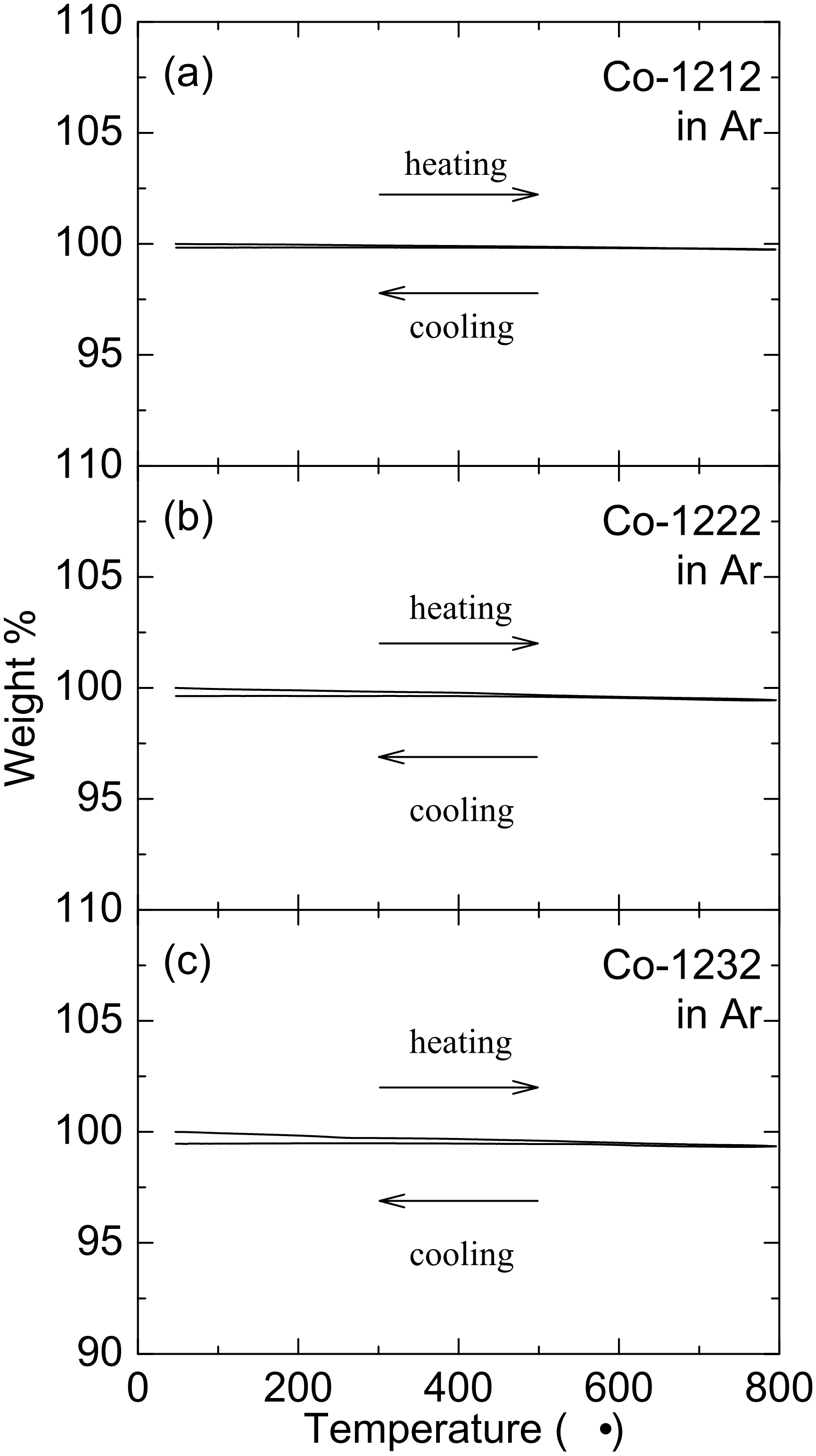

PCP49 / Figure 3 (Morita *et al.*)

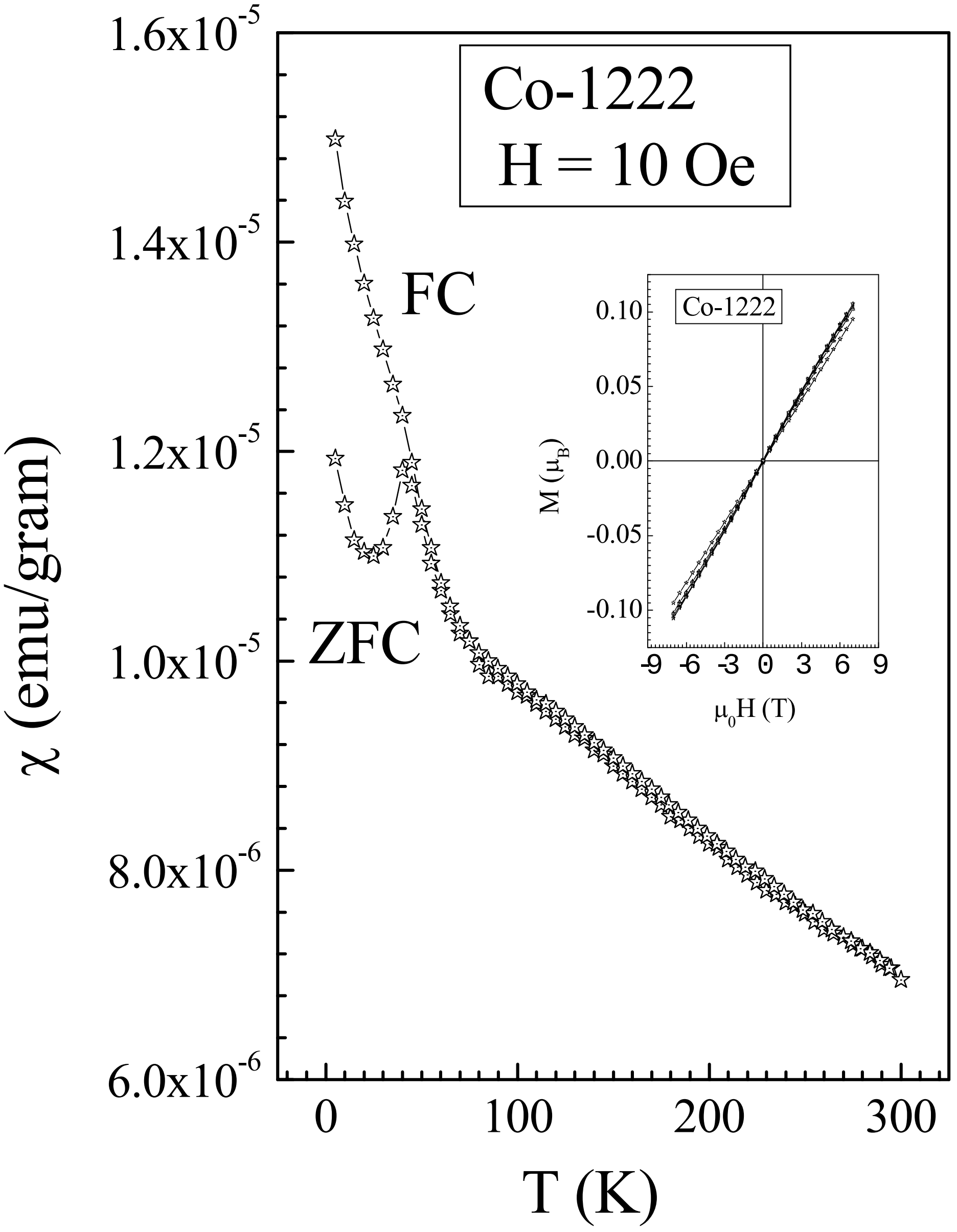